# An assessment of higher gradient theories from a continuum mechanics perspective

Ali R. Hadjesfandiari, Gary F. Dargush

*Department of Mechanical and Aerospace Engineering*
*University at Buffalo, The State University of New York, Buffalo, NY 14260 USA*

ah@buffalo.edu, gdargush@buffalo.edu

October 11, 2018

**Abstract**

In this paper, we investigate the inherent physical and mathematical character of higher gradient theories, in which the strain or distortion gradients are considered as the fundamental measures of deformation. Contrary to common belief, the first or higher strain or distortion gradients are not proper measures of deformation. Consequently, their corresponding energetically conjugate stresses are non-physical and cannot represent the state of internal stresses in the continuum. Furthermore, the governing equations in these theories do not describe the motion of infinitesimal elements of matter consistently. For example, in first strain gradient theory, there are nine governing equations of motion for infinitesimal elements of matter at each point; three force equations, and six unsubstantiated artificial moment equations that violate Newton's third law of action and reaction and the angular momentum theorem. This shows that the first strain gradient theory (F-SGT) is not an extension of rigid body mechanics, which then is not recovered in the absence of deformation. The inconsistencies of F-SGT and other higher gradient theories also manifest themselves in the appearance of strains, distortions or their gradients as boundary conditions and the requirement for many material coefficients in the constitutive relations.

# 1. Introduction

Mindlin and Tiersten (1962) and Koiter (1964) developed the initial version of couple stress theory (MTK-CST), in which the deformation is completely specified by the continuous displacement field. This theory is implicitly based on the rigid body portion of motion of infinitesimal elements of matter at each point of the continuum (Hadjesfandiari and Dargush, 2011, 2015b). Therefore, in this theory, the internal stresses are exactly the force- and couple-stress tensors, introduced by Cosserat and Cosserat (1909), each having at most nine independent components. However, MTK-CST suffers from some serious inconsistencies and difficulties with the underlying formulations (Hadjesfandiari and Dargush, 2015a,b). Much earlier, for example, Eringen (1968) realized the indeterminacy of the couple-stress tensor as a major mathematical problem in the original MTK-CST, which he afterwards called indeterminate couple stress theory. Instead of resolving the inconsistencies of MTK-CST in the framework of this theory, researchers turned to other theories, such as micropolar theories (Eringen (1968) and higher gradient theories, such as strain gradient theories (Mindlin, 1965; Mindlin and Eshel, 1968).

Remarkably, the inconsistencies of MTK-CST have been resolved by discovering the subtle skew-symmetric character of the couple-stress tensor (Hadjesfandiari and Dargush, 2011; Hadjesfandiari et al., 2015). Therefore, in this theory, called consistent couple stress theory (C-CST), the couple-stress tensor is determinate and creates bending deformation. Consequently, C-CST resolves the quest for a consistent size-dependent continuum mechanics by answering the criticism of Eringen, and provides a fundamental basis for the development of size-dependent material response.

In gradient theories, such as strain gradient theories (SGT) and distortion gradient theories (DGT), the indeterminacy noted above is apparently avoided by taking strain or distortion gradients as measures of deformation. However, the state of internal stresses dramatically deviates from that defined in the Cosserat and couple stress theories. As will be shown, the new stresses in gradient theories have no relation with the concept of the classical moment of forces, which indicates that these theories suffer from new physical and mathematical inconsistencies. For example, in first strain gradient theory (F-SGT) (Mindlin and Eshel, 1968; Georgiadis et al., 2000; Aifantis and Willis, 2006), the strain gradient with 18 components is taken as the measure of deformation, which in turn requires a new third order double-stress with 18 components. However, this non-physical third order double-stress tensor is different from the physical couple-stress tensor, appearing in Cosserat and couple stress theories, and makes the force-stress tensor in F-SGT symmetric. More surprisingly, there are nine governing equations of motion for an infinitesimal element of matter in F-SGT; three force equations and six non-physical first symmetric moment equations.

We should also notice that not all of the higher order continuum mechanics theories could be correct at the same time, because they can predict contradicting results. For example, couple stress theory (CST) predicts no size-effect for a pressurized long thick-walled isotropic elastic cylinder. This is because the axisymmetric pressure loading creates a radial deformation without any rotation and curvature. On the other hand, the strain gradient theories (SGT) predict size-effect for this problem. Gao and Park (2007) have shown this size-effect based on a simplified first strain gradient elasticity theory suggested by Altan and Aifantis (1997), which requires only one strain gradient elastic coefficient. However, this simplified F-SGT does not describe bending of beams and plates properly, which in hindsight indicates that there is a fundamental inconsistency in this simplified strain gradient theory, and thus, in the more general strain gradient theories.

The remainder of the paper is organized as follows. In Section 2, we briefly examine the fundamental character of the consistent governing equations of motion. This includes a concise presentation of couple stress theory. Afterward, in Section 3, we develop continuum mechanics based on the first symmetric moment equation. In Section 4, we demonstrate that F-SGT resembles this inconsistent theory, where there are nine governing equations of motion for infinitesimal elements of matter at each point, thus violating Newton's third law of action and reaction and the angular momentum theorem. Subsequently, in Section 5, we briefly examine the applications of F-SGT in elasticity, plasticity, size-dependent piezoelectricity and flexoelectricity. Section 6 provides a discussion on higher gradient theory, including second gradient theories. Finally, Section 7 contains a summary and some general conclusions.

## 2. Fundamental governing equations of motion for continua

Consider a material continuum occupying a volume $V$ bounded by a surface $S$. The deformation of the body is represented by the continuous displacement field $u_i$. In infinitesimal deformation theory, the displacement vector field $u_i$ is sufficiently small that the infinitesimal strain and rotation tensors are defined as

$$e_{ij} = u_{(i,j)} = \frac{1}{2}\left(u_{i,j} + u_{j,i}\right) \tag{1}$$

$$\omega_{ij} = u_{[i,j]} = \frac{1}{2}\left(u_{i,j} - u_{j,i}\right) \tag{2}$$

respectively. Here standard indicial notation is used and we have introduced parentheses to denote the symmetric part of a second order tensor, whereas square brackets are associated with the skew-symmetric part. Since the true (polar) rotation tensor $\omega_{ij}$ is skew-symmetrical, one can introduce its corresponding dual pseudo (axial) rotation vector as

$$\omega_i = \frac{1}{2}\varepsilon_{ijk}\omega_{kj} = \frac{1}{2}\varepsilon_{ijk}u_{k,j} \tag{3}$$

where $\varepsilon_{ijk}$ represents the permutation or Levi-Civita symbol.

In consistent continuum mechanics, we consider the rigid body portion of motion of infinitesimal elements of matter (or rigid triads) at each point of the continuum (Hadjesfandiari and Dargush, 2015b). Therefore, the degrees of freedom are the displacements $u_i$ and rotations $\omega_i$ at each point, which describe, respectively, the translation and rotation of an infinitesimal element of matter in the neighborhood of the point. However, the continuity of matter within the continuum description restrains the rotation field $\omega_i$ by the relation (3). This of course shows that the rotation field $\omega_i$ in the interior of $V$ is not independent of the displacement field $u_i$.

For the continuum at any point $x_i$, the velocity and acceleration fields are $v_i = \dfrac{Dx_i}{Dt}$ and $a_i = \dfrac{Dv_i}{Dt}$, where $\dfrac{D}{Dt}$ is the material or substantial derivative. However, in small deformation theory, we can use the approximation $v_i = \dfrac{\partial u_i}{\partial t} = \dot{u}_i$ and $a_i = \dfrac{\partial^2 u_i}{\partial t^2} = \ddot{u}_i$.

The governing equations must describe the motion of infinitesimal elements of matter at each point. The fundamental governing equations in consistent continuum mechanics are based on the force and moment equations for a system of particles (Hadjesfandiari and Dargush, 2018)

$$\sum F_i^{ext} = \sum ma_i \qquad \sum \mathbf{F}^{ext} = \sum m\mathbf{a} \qquad (4)$$

$$\sum M_i^{ext} = \varepsilon_{ijk} \sum x_j ma_k \qquad \sum \mathbf{M}^{ext} = \sum \mathbf{r} \times m\mathbf{a} \qquad (5)$$

where $\sum F_i^{ext}$ and $\sum M_i^{ext}$ are the total external forces and external moments exerted on the system, respectively, and $ma_i$ is the time rate of change of linear momentum of each particle with mass $m$ and acceleration $a_i$. We notice that the fundamental governing equations (4) and (5) are the result of Newton's second and third laws for the system of particles.

In the corresponding continuum mechanics, called couple stress theory (CST) (Mindlin and Tiersten,1962; Koiter, 1964; Hadjesfandiari and Dargush, 2011, 2015b), the transfer of the interaction in the continuum through a surface element $dS$ with unit normal vector $n_i$ occurs by means of a force vector $t_i^{(n)}dS$ and a couple vector $m_i^{(n)}dS$, where $t_i^{(n)}$ and $m_i^{(n)}$ are the force-traction vector and couple-traction vector, respectively. Therefore, in this couple stress theory, the internal stresses are represented by the second order true (polar) force-stress tensor $\sigma_{ij}$ and the second order pseudo (axial) couple-stress tensor $\mu_{ij}$, where

$$t_i^{(n)} = \sigma_{ji} n_j \qquad (6)$$

$$m_i^{(n)} = \mu_{ji} n_j \qquad (7)$$

For a continuum, equations (4) and (5) lead to the governing equations in couple stress theory (Mindlin and Tiersten,1962; Koiter, 1964; Hadjesfandiari and Dargush, 2011, 2015b) as

$$u_i \rightarrow 3 \text{ force equations} \qquad \sigma_{ji,j} + f_i = \rho a_i \qquad (8)$$

$$\omega_i \rightarrow 3 \text{ moment equations} \qquad \mu_{ji,j} + \varepsilon_{ijk}\sigma_{jk} = 0 \qquad (9)$$

which describe the translational and rotational motion of an element of matter corresponding to $u_i$ and $\omega_i$, respectively. Here $f_i$ is the specified body-force density and $\rho$ is the mass density.

Hadjesfandiari and Dargush (2011, 15b, 2018) have developed consistent couple stress theory (C-CST) by discovering, not merely assuming, the skew-symmetric character of the couple stress tensor, that is

$$\mu_{ji} = -\mu_{ij} \qquad (10)$$

The fundamental governing equations for a system of particles can also be written in the form of force and skew-symmetric moment equations (Hadjesfandiari and Dargush, 2018)

$$\sum F_i^{ext} = \sum ma_i \qquad (11)$$

$$\sum M_{ij}^{ext} = \sum \frac{1}{2}\left(ma_i x_j - ma_j x_i\right) \qquad (12)$$

where $\sum M_{ij}^{ext}$ is the true (polar) skew-symmetric external moment tensor dual to $\sum M_i^{ext}$. This form is more illuminating for the analysis to follow.

In couple stress theory based on the true tensor form (Hadjesfandiari and Dargush, 2018), the skew-symmetric couple-traction vector $m_{ij}^{(n)}$ is the dual of the couple-traction vector $m_i^{(n)}$, where

$$m_{ij}^{(n)} = \frac{1}{2}\varepsilon_{jik}m_k^{(n)}, \qquad m_i^{(n)} = \varepsilon_{ijk}m_{kj}^{(n)} \tag{13}$$

Consequently, the governing equations for an infinitesimal element of matter can be written as (Hadjesfandiari and Dargush, 2018)

$$\sigma_{ji,j} + f_i = \rho a_i \tag{14}$$

$$\sigma_{[ij]} + \mu_{kij,k} = 0 \tag{15}$$

where the second order pseudo couple-stress tensor $\mu_{ij}$ has been replaced with the third order true couple-stress $\mu_{ijk}$ tensor, for which

$$\mu_{ikj} = -\mu_{ijk} \tag{16}$$

For more details, see Hadjesfandiari and Dargush (2018).

## 3. Continuum mechanics based on the symmetric moment equation

Now, we investigate the character of continuum mechanics, when the fundamental governing equations for a system of particles are based on the force and the invalid and non-physical first symmetric moment equations (Hadjesfandiari and Dargush, 2018). For this case,

$$\sum F_i^{ext} = \sum ma_i \tag{17}$$

$$\sum \tilde{M}_{ij}^{ext} = \sum \frac{1}{2}(ma_i x_j + ma_j x_i) \tag{18}$$

Here $\sum \tilde{M}_{ij}^{ext}$ is the total external symmetric moment of external forces exerted on the system, where the first symmetric moment of a force $F_i$ at $x_i$ about the origin is given by the tensor $\tilde{M}_{ij} = \frac{1}{2}(F_i x_j + F_j x_i)$. We notice that the first symmetric moment of internal forces $\sum \tilde{M}_{ij}^{int}$ has been ignored in writing (18), which violates Newton's third law of action and reaction (Hadjesfandiari and Dargush, 2018) and is the reason that this relation is crossed out. We also notice that the symmetric moment tensor of a couple consisting of $F_i$ and $-F_i$ depends on the position of origin, and cannot describe its physical effect properly.

In the corresponding continuum mechanics, it is postulated that the transfer of the interaction in the continuum through a surface element $dS$ with unit normal vector $n_i$ occurs by means of a force vector $t_i^{(n)} dS$ and a doublet (or couple) with symmetric moment tensor $\tilde{m}_{ij}^{(n)} dS$, where

$$\tilde{m}_{ji}^{(n)} = \tilde{m}_{ij}^{(n)} \qquad (19)$$

Here $t_i^{(n)}$ and $\tilde{m}_{ij}^{(n)}$ are the force-traction vector and first symmetric double-traction tensor, respectively.

The state of internal stress at each point is known, if the force-traction vector $t_i^{(n)}$ and first symmetric double-traction tensor $\tilde{m}_{ij}^{(n)}$ on arbitrary surfaces at that point are known. This requires knowledge of only the force-traction and symmetric double-traction on three mutually independent planes passing the point. When these planes are taken parallel to the coordinate planes with unit normal $n_i$ along the coordinate axes, the force-traction vectors are $t_i^{(1)}$, $t_i^{(2)}$ and $t_i^{(3)}$, and the first symmetric double-traction tensors are $\tilde{m}_{ij}^{(1)}$, $\tilde{m}_{ij}^{(2)}$ and $\tilde{m}_{ij}^{(3)}$. Consequently, in this continuum theory, the internal stresses are represented by the second order force-stress tensor $\sigma_{ij}$ and the third order symmetric true double-stress tensor $\tilde{\mu}_{ijk}$ with the symmetry condition

$$\tilde{\mu}_{ijk} = \tilde{\mu}_{ikj} \tag{20}$$

where we have the relations

$$\sigma_{ij} = t_j^{(i)} \tag{21}$$

$$\tilde{\mu}_{ijk} = \tilde{m}_{jk}^{(i)} \tag{22}$$

We notice that the tensors $\sigma_{ij}$ and $\tilde{\mu}_{ijk}$ can have up to 9 and 18 independent components, respectively. Accordingly, the force-traction vector $t_i^{(n)}$ and double-traction vector $\tilde{m}_{ij}^{(n)}$ are given as

$$t_i^{(n)} = \sigma_{ji} n_j \tag{23}$$

$$\tilde{m}_{ij}^{(n)} = \tilde{\mu}_{kij} n_k \tag{24}$$

Since the first symmetric moment does not describe the physical effect of a couple consisting of $F_i$ and $-F_i$ correctly, the symmetric double-traction $\tilde{m}_{ij}^{(n)}$ tensor and the double-stress tensor $\tilde{\mu}_{ijk}$ do not describe properly the effect of physical couple-traction $m_i^{(n)}$ (or $m_{ij}^{(n)}$) and couple-stress $\mu_{ij}$ (or $\mu_{ijk}$) as introduced by Cosserat and Cosserat (1909).

To obtain the governing equations in this theory, we apply the force and the first symmetric moment equations (17) and (18) for an arbitrary part of the material continuum occupying a volume $V_a$ enclosed by boundary surface $S_a$. Accordingly, we have

$$\int_{S_a} t_i^{(n)} dS + \int_{V_a} f_i dV = \int_{V_a} \rho a_i dV \tag{25}$$

$$\int_{S_a}\left[\frac{1}{2}\left(x_j t_i^{(n)} + x_i t_j^{(n)}\right) + \tilde{m}_{ij}^{(n)}\right] dS + \int_{V_a} \frac{1}{2}\left(x_j f_i + x_i f_j\right) dV = \int_{V_a}\left[\frac{1}{2}\left(x_j \rho a_i + x_i \rho a_j\right)\right] dV \quad (26)$$

By using the relations (23) and (24) for tractions, the force and first symmetric moment governing equation (25) and (26) can be written as

$$\int_{S_a} \sigma_{ji} n_j dS + \int_{V_a} f_i dV = \int_{V_a} \rho a_i dV \quad (27)$$

$$\int_{S_a}\left[\frac{1}{2}\left(x_j \sigma_{ki} n_k + x_i \sigma_{kj} n_k\right) + \tilde{\mu}_{kij} n_k\right] dS + \int_{V_a} \frac{1}{2}\left(x_j f_i + x_i f_j\right) dV$$
$$= \int_{V_a}\left[\frac{1}{2}\left(x_j \rho a_i + x_i \rho a_j\right)\right] dV \quad (28)$$

Now we use the divergence theorem to obtain

$$\int_{V_a}\left[\sigma_{ji,j} + f_i\right] dV = \int_{V_a} \rho a_i dV \quad (29)$$

$$\int_{V_a}\left[\frac{1}{2}\left(\delta_{jk}\sigma_{ki} + x_j \sigma_{ki,k} + \delta_{ik}\sigma_{kj} + x_i \sigma_{kj,k}\right) + \tilde{\mu}_{kij,k}\right] dV + \int_{V_a} \frac{1}{2}\left(x_j f_i + x_i f_j\right) dV$$
$$= \int_{V_a}\left[\frac{1}{2}\left(x_j \rho a_i + x_i \rho a_j\right)\right] dV \quad (30)$$

After considering the arbitrariness of volume $V_a$ in (29), we obtain the differential form of the force equation as

$$\sigma_{ji,j} + f_i = \rho a_i \quad (31)$$

The first symmetric moment governing equation (30) can also be written as

$$\int_{V_a}\left[\frac{1}{2}\left(\sigma_{ji} + \sigma_{ij}\right) + \frac{1}{2}x_j\left(\sigma_{ki,k} + f_i - \rho a_i\right) + \frac{1}{2}x_i\left(\sigma_{kj,k} + f_j - \rho a_j\right) + \tilde{\mu}_{kij,k}\right] dV = 0 \quad (32)$$

Then, by using the force governing equation (31), this reduces to

$$\int_{V_a}\left[\frac{1}{2}\left(\sigma_{ij} + \sigma_{ji}\right) + \tilde{\mu}_{kij,k}\right] dV = 0 \quad (33)$$

By noticing the arbitrariness of volume $V_a$, we finally obtain the differential form of the first symmetric moment equation as

$$\frac{1}{2}\left(\sigma_{ij} + \sigma_{ji}\right) + \tilde{\mu}_{kij,k} = 0 \tag{34}$$

Therefore, in this non-physical inconsistent continuum mechanics, the governing equations for an infinitesimal element of matter are (31) and (34) rewritten as

$$\sigma_{ji,j} + f_i = \rho a_i \tag{35}$$

$$\sigma_{(ji)} + \tilde{\mu}_{kij,k} = 0 \tag{36}$$

The number of governing equations of motion in this theory is nine. The three force equations (35) correspond to the translational motions $u_i$, while the six symmetric moment equations (36) correspond to unknown fictitious degrees of freedom $\varpi_{ij}$ with the symmetry relation

$$\varpi_{ji} = \varpi_{ij} \tag{37}$$

Thus,

$$u_i \rightarrow 3 \text{ force equations} \qquad \sigma_{ji,j} + f_i = \rho \ddot{u}_i \tag{38}$$

$$\varpi_{ij} \rightarrow 6 \text{ double-force equations} \qquad \sigma_{(ji)} + \tilde{\mu}_{kij,k} = 0 \tag{39}$$

This result contradicts the fact that $u_i$ and $\omega_i$ are the degrees of freedom describing the motion of an infinitesimal element of matter. We notice that the physical moment equation (9) (or its equivalent (15)) has been replaced by the completely non-physical symmetric first moment equations (36). Instead of producing the skew-symmetric part of the force-stress tensor, the non-physical first symmetric moment equation (36) gives the symmetric part of the force-stress tensor as

$$\sigma_{(ji)} = -\tilde{\mu}_{kij,k} \tag{40}$$

Thus, for the total force-stress tensor, we obtain

$$\sigma_{ji} = \sigma_{(ji)} + \sigma_{[ji]} \qquad (41)$$
$$= -\tilde{\mu}_{kij,k} + \sigma_{[ji]}$$

By using this form of the total force-stress tensor, the linear governing equation of motion (35) reduces to

$$\left[\sigma_{[ji]} - \tilde{\mu}_{kij,k}\right]_{,j} + f_i = \rho a_i \qquad (42)$$

As we can see, this governing equation is only a set of three equations representing the force or linear equation of motion, which can be called the reduced linear governing equation in this inconsistent theory. Since (42) is a combination of the original force and moment equations (35) and (36), it cannot be considered as a fundamental law by itself. This can be confirmed by noticing that the highest derivative in this governing equation is of second order.

Surprisingly, it turns out that the first strain gradient theory (F-SGT) (Mindlin and Eshel, 1968; Georgiadis et al., 2000; Aifantis and Willis, 2006) resembles this non-physical continuum mechanics theory. We investigate the character of this popular first strain gradient theory (F-SGT) in more detail in the next section.

Interestingly, when the effect of double-stresses $\tilde{\mu}_{ijk}$ are neglected, the governing equations (35) and (36) reduce to

$$\sigma_{ji,j} + f_i = \rho a_i \qquad (43)$$
$$\sigma_{(ij)} = 0 \qquad (44)$$

which also can be written

$$\sigma_{[ji],j} + f_i = \rho a_i \qquad (45)$$
$$\sigma_{ji} = \sigma_{[ji]} \qquad (46)$$

Thus, in the absence of double-stresses $\tilde{\mu}_{ijk}$, F-SGT reduces to a non-physical theory with skew-symmetric force-stress tensor.

## 4. First strain gradient theory

General higher gradient theories were introduced in the 1960s by Kröner (1963), Kröner and Datta (1966), Kröner (1967), Mindlin (1964, 1965), Mindlin and Eshel (1968), Green and Rivlin (1964a,b). In these theories, various forms of the gradient of displacement (distortion) or gradient of the strain tensor, such as $u_{i,j}$, $u_{i,jk}$, $u_{i,jkl}$, $e_{ij,k}$ or $e_{ij,kl}$ have been taken as fundamental measures of deformation. Here, we concentrate on the more popular case, where the gradient of strain $e_{ij,k}$ is a measure of deformation. However, we also briefly examine the character of higher gradient theories with other measures of deformation in the following sections.

In the simplest higher strain gradient theory, called first strain gradient theory (F-SGT) (Mindlin and Eshel, 1968; Georgiadis et al., 2000; Aifantis and Willis, 2006), the strain gradient tensor

$$\tilde{k}_{ijk} = e_{ij,k} \tag{47}$$

is taken as the higher measure of deformation with the obvious symmetry

$$\tilde{k}_{ijk} = \tilde{k}_{jik} \tag{48}$$

However, this contradicts our notion in continuum mechanics that the measures of deformation are defined based on the gradient of degrees of freedom of an infinitesimal element of matter at each point (Hadjesfandiari and Dargush, 2015b). For example, the strain tensor $e_{ij}$ as a measure of deformation is defined from the displacement vector $u_i$, which are the translational degrees of freedom. However, the third order true strain gradient tensor $\tilde{k}_{ijk}$ with 18 independent components is not a suitable measure or metric of deformation, because it is not defined directly from the gradient of the degrees of freedom $u_i$ or $\omega_i$.

By taking $e_{ij}$ and $\tilde{k}_{ijk}$ as measures of deformation in F-SGT, the theory implicitly considers $u_i$ and $\varpi_{ij} = e_{ij}$ as the degrees of freedom, where

$$u_i \quad \to \quad e_{ij} = \frac{1}{2}\left(u_{i,j} + u_{j,i}\right) \tag{49}$$

$$e_{ij} \quad \to \quad \tilde{k}_{ijk} = e_{ij,k} \tag{50}$$

Therefore, in F-SGT, the strain tensor $e_{ij}$ is a measure of deformation and a set of degrees of freedom at the same time. Consequently, in F-SGT, the number of degrees of freedom for infinitesimal elements of matter is nine; three corresponding to translations $u_i$ and six corresponding to the fictitious degrees of freedom $\varpi_{ij} = e_{ij}$. This in turn requires that the number of governing equations of motion is nine; three corresponding to translations $u_i$ and six corresponding to strain degrees of freedom $e_{ij}$. This result clearly shows that F-SGT is not based on considering the motion of infinitesimal elements of matter at each point properly, where the degrees of freedom are $u_i$ and $\omega_i$ (or $\omega_{ij}$).

Since the third order true strain gradient tensor $\tilde{k}_{ijk}$ is a measure of deformation in F-SGT, it requires a third order symmetric true double-stress tensor with 18 components similar to the previously defined $\tilde{\mu}_{ijk}$ in Section 3 with the symmetry relation

$$\tilde{\mu}_{ijk} = \tilde{\mu}_{ikj} \tag{51}$$

as the corresponding energetically conjugate stress tensor. Therefore, in F-SGT, the internal stresses are represented by the force-traction vector $\tilde{t}_i^{(n)}$ and first symmetric double-traction tensor $\tilde{m}_{ij}^{(n)}$, where

$$\tilde{m}_{ji}^{(n)} = \tilde{m}_{ij}^{(n)} \tag{52}$$

In this formulation, the force-stress tensor $\sigma_{ij}$ is symmetric and can be decomposed into two symmetric parts $\sigma'_{ij}$ and $\sigma''_{ij}$, where

$$\sigma_{ij} = \sigma'_{ij} + \sigma''_{ij} = \sigma_{ji} \tag{53}$$

Here $\sigma'_{ij}$ is the symmetric force-stress tensor corresponding to its counterpart in classical theory, while $\sigma''_{ij}$ represents the symmetric force-stress tensor corresponding to the effect of $\tilde{\mu}_{ijk}$. In this F-SGT, the governing equations are given as (Mindlin and Eshel, 1968; Georgiadis et al., 2000; Aifantis and Willis, 2006)

$$\sigma_{ji,j} + f_i = \rho a_i \tag{54}$$

$$\sigma''_{(ji)} + \tilde{\mu}_{kij,k} = 0 \tag{55}$$

These specifications show that F-SGT resembles the non-physical continuum mechanics developed in Section 3. Notice that the set of governing equations (54) and (55) are similar to the set of governing equations (35) and (36), which are based on the force equation (17) and the non-physical first symmetric moment equation (18) for a system of particles, respectively. However, we note that the governing equations of motion (54) and (55) in F-SGT are not usually obtained from the force and non-physical first symmetric moment equations (17) and (18). Instead, these inconsistent governing equations are obtained by using variational methods or the virtual work principle. This is the reason why the researchers in F-SGT do not usually realize the existence of so many inconsistencies in this theory, such as the non-physical character of the symmetric double-traction tensor $\tilde{m}_{ij}^{(n)}$ and double-stress tensor $\tilde{\mu}_{ijk}$, and the existence of nine non-physical governing equations of motion (54) and (55). The symmetric double-traction $\tilde{m}_{ij}^{(n)}$ in F-SGT is based on the non-physical symmetric moment tensor $\tilde{M}_{ij} = \frac{1}{2}(F_i x_j + F_j x_i)$, rather than on the physical skew-symmetric moment $M_{ij} = \frac{1}{2}(F_i x_j - F_j x_i) = -M_{ji}$ or its dual pseudo-vector $M_i = \varepsilon_{ijk} x_j F_k$ ($\mathbf{M} = \mathbf{r} \times \mathbf{F}$).

We investigate the character of the popular F-SGT in more detail by using a virtual work principle to derive the governing equations and boundary conditions as follows.

## 4.1. Consequence of the virtual work principle for first strain gradient theory

We notice that in F-SGT the tractions $t_i^{(n)}$ and $\tilde{m}_{ij}^{(n)}$ are energetically conjugate to $u_i$ and $\varpi_{ij} = e_{ij}$, respectively, that is

$$u_i \leftrightarrow t_i^{(n)} \tag{56}$$

$$e_{ij} \leftrightarrow \tilde{m}_{ij}^{(n)} \tag{57}$$

Therefore, the virtual work of the force $t_i^{(n)} dS$ and the symmetric double $\tilde{m}_{ij}^{(n)} dS$ system on the surface element $dS$ is $\left( t_i^{(n)} \delta u_i + \tilde{m}_{ij}^{(n)} \delta e_{ij} \right) dS$.

For F-SGT, the virtual work principle is

$$\delta W_{ext} = \delta W_{int} + \int_V \rho a_i \delta u_i dV \tag{58}$$

where the external virtual work is

$$\delta W_{ext} = \int_S \left[ t_i^{(n)} \delta u_i + \tilde{m}_{ij}^{(n)} \delta e_{ij} \right] dS + \int_V f_i \delta u_i dV \tag{59}$$

However, in this theory the internal virtual work is forced to be

$$\delta W_{int} = \int_V \left[ \sigma'_{ji} \delta e_{ij} + \tilde{\mu}_{kij} \delta \tilde{k}_{ijk} \right] dV \tag{60}$$

Therefore, the virtual work theorem for this formulation is

$$\int_S \left[ t_i^{(n)} \delta u_i + \tilde{m}_{ij}^{(n)} \delta e_{ij} \right] dS + \int_V f_i \delta u_i dV = \int_V \left[ \sigma'_{ji} \delta e_{ij} + \tilde{\mu}_{kij} \delta \tilde{k}_{ijk} \right] dV + \int_V \rho a_i \delta u_i dV \tag{61}$$

We notice that the strain tensor $e_{ij}$ is a set of degrees of freedom in the external virtual work (59), and a measure of deformation in the internal virtual work (60).

As mentioned previously, in this formulation the force-stress tensor $\sigma_{ij}$ is symmetric and has been decomposed into two symmetric parts $\sigma'_{ij}$ and $\sigma''_{ij}$, where

$$\sigma_{ij} = \sigma'_{ij} + \sigma''_{ij} = \sigma_{ji} \tag{62}$$

As the virtual work principle for F-SGT shows, the force-stress tensor $\sigma'_{ij}$ and the double-stress tensor $\tilde{\mu}_{ijk}$ are energetically conjugate to the strain tensor $e_{ij}$ and the strain gradient tensor $\tilde{k}_{ijk} = e_{ij,k}$, respectively, where

$$\sigma'_{ij} \leftrightarrow e_{ij} \tag{63}$$

$$\tilde{\mu}_{kij} \leftrightarrow \tilde{k}_{ijk} = e_{ij,k} \tag{64}$$

Surprisingly, the virtual work principle (61) requires that the symmetric force-stress part $\sigma''_{ij}$ be workless, whereas the symmetric force-stress part $\sigma'_{ij}$ and double-stress tensor $\tilde{\mu}_{ijk}$ do work. This strange character clearly is the result of the inconsistency of F-SGT.

Depending on how the kinematical constraint

$$e_{ij} - \frac{1}{2}\left(u_{i,j} + u_{j,i}\right) = 0 \tag{65}$$

is enforced in the virtual work principle (61), we can continue with either of the following alternative methods:

1. We directly enforce the kinematical constraint (65) in the virtual work principle (61), which results in one set of reduced governing equations corresponding to the degrees of freedom $u_i$.

2. We enforce the kinematical constraint (65) in the virtual work principle (61) by using the Lagrange multiplier method. This makes the independent degrees of freedoms $u_i$ and $\varpi_{ij} = e_{ij}$ also independent variables in the virtual work principle. This approach results in two sets of fundamental governing equations corresponding to the degrees of freedom $u_i$ and $e_{ij}$.

We demonstrate the details of these methods as follows.

*4.1.1. Direct method*

The internal virtual work $\delta W_{int}$ (60) can be written as

$$\delta W_{int} = \int_V \left[ \sigma'_{(ji)} \delta e_{ij} + \tilde{\mu}_{kij} \delta e_{ij,k} \right] dV$$

$$= \int_V \left[ \sigma'_{(ji)} \delta e_{ij} + \left( \tilde{\mu}_{kij} \delta e_{ij} \right)_{,k} - \tilde{\mu}_{kij,k} \delta e_{ij} \right] dV \qquad (66)$$

$$= \int_V \left\{ \left[ \sigma'_{(ji)} - \tilde{\mu}_{kij,k} \right] \delta e_{ij} + \left( \tilde{\mu}_{kij} \delta e_{ij} \right)_{,k} \right\} dV$$

Then, by some manipulation, we obtain

$$\delta W_{int} = \int_V \left\{ \left( \left[ \sigma'_{(ji)} - \tilde{\mu}_{kij,k} \right] \delta u_i \right)_{,j} + \left( \tilde{\mu}_{kij} \delta e_{ij} \right)_{,k} - \left[ \sigma'_{(ji)} - \tilde{\mu}_{kij,k} \right]_{,j} \delta u_i \right\} dV \qquad (67)$$

By using the divergence theorem, this can be rewritten as

$$\delta W_{int} = \int_S \left\{ \left[ \sigma'_{(ji)} - \tilde{\mu}_{kij,k} \right] n_j \delta u_i + \tilde{\mu}_{kij} n_k \delta e_{ij} \right\} dS - \int_V \left[ \sigma_{(ji)} - \tilde{\mu}_{kij,k} \right]_{,j} \delta u_i dV \qquad (68)$$

Therefore, the virtual work principle (61) becomes

$$\int_S \left[ t_i^{(n)} \delta u_i + \tilde{m}_{ij}^{(n)} \delta e_{ij} \right] dS + \int_V f_i \delta u_i dV$$

$$= \int_S \left\{ \left[ \sigma'_{(ji)} - \tilde{\mu}_{kij,k} \right] n_j \delta u_i + \tilde{\mu}_{kij} n_k \delta e_{ij} \right\} dS - \int_V \left[ \sigma_{(ji)} - \tilde{\mu}_{kij,k} \right]_{,j} \delta u_i dV + \int_V \rho a_i \delta u_i dV \qquad (69)$$

Now by noticing that the variation of $\delta u_i$ is arbitrary in the domain, we obtain only one set of governing equations of motion corresponding to the degrees of freedom $u_i$, which is

$$\left[ \sigma'_{(ji)} - \tilde{\mu}_{kij,k} \right]_{,j} + f_i = \rho a_i \qquad (70)$$

and the tractions as

$$t_i^{(n)} = \left[ \sigma'_{(ji)} - \tilde{\mu}_{kij,k} \right] n_j \qquad (71)$$

$$\tilde{m}_{ij}^{(n)} = \tilde{\mu}_{kij} n_k \qquad (72)$$

It should be noted that the governing equation (70) is a set of only three equations representing the linear equilibrium equations. There is no explicit first symmetric moment equation corresponding to the degrees of freedom $e_{ij}$. However, we notice that the governing equation (70) is a combination of the original force and the symmetric moment-like equations

$$\sigma_{ji,j} + f_i = \rho a_i \tag{73}$$

$$\sigma''_{ij} = -\tilde{\mu}_{kij,k} \tag{74}$$

It is obvious that the set of six moment-like equations (74) corresponds to the degrees of freedom $e_{ij}$.

In some developments based on F-SGT, the reduced governing equation (70) has been given as the sole governing equation (Huang et al., 2000; Jiang et al., 2001; Hwang et al., 2002; Bažant and Guo, 2002; Qiu et al., 2003; Zhao and Pedroso, 2008; Lim et al., 2015; Zhou et al., 2016; Gourgiotis et al., 2018). However, this reduced governing equation cannot be considered as a fundamental law by itself. We must not forget about the six symmetric moment equations (74), which have been combined with three force equations to produce (70). This is the reason why the proponents of F-SGT are not usually aware of the fact that the displacements $u_i$ and the strains $e_{ij}$ are the primary degrees of freedom, and there are nine governing equations of motion inherent to this theory.

### 4.1.2. Lagrange multiplier method

By using the Lagrange multiplier method, we transform the virtual work principle (58) to

$$\delta W_{ext} = \delta W^*_{int} + \int_V \rho a_i \, \delta u_i \, dV \tag{75}$$

where the augmented internal virtual work $\delta W^*_{int}$ is

$$\delta W^*_{int} = \int_V \left\{ \sigma'_{(ji)} \delta u_{i,j} + \tilde{\mu}_{kij} \delta e_{ij,k} - \lambda_{ij} \left[ \delta e_{ij} - \frac{1}{2}\left( \delta u_{i,j} + \delta u_{j,i} \right) \right] \right\} dV \tag{76}$$

We notice that the Lagrange multiplier tensor $\lambda_{ij}$ is symmetric

$$\lambda_{ji} = \lambda_{ij} \tag{77}$$

The augmented internal virtual work $\delta W_{int}^*$ can be written as

$$\delta W_{int}^* = \int_V \left\{ \left[ \sigma'_{(ji)} + \lambda_{ij} \right] \delta u_{i,j} + \left( \tilde{\mu}_{kij} \delta e_{ij} \right)_{,k} - \left( \tilde{\mu}_{kij,k} + \lambda_{ij} \right) \delta e_{ij} \right\} dV \tag{78}$$

or

$$\delta W_{int}^* = \int_V \left\{ \left( \left[ \sigma'_{(ji)} + \lambda_{ij} \right] \delta u_i \right)_{,j} - \left[ \sigma'_{(ji)} + \lambda_{ij} \right]_{,j} \delta u_i + \left( \tilde{\mu}_{kij} \delta e_{ij} \right)_{,k} - \left( \tilde{\mu}_{kij,k} + \lambda_{ij} \right) \delta e_{ij} \right\} dV \tag{79}$$

By using the divergence theorem, we obtain

$$\delta W_{int}^* =$$
$$= \int_S \left\{ \left[ \sigma'_{(ji)} + \lambda_{ij} \right] n_j \delta u_i + \tilde{\mu}_{kij} n_k \delta e_{ij} \right\} dS - \int_V \left\{ \left[ \sigma'_{(ji)} + \lambda_{ij} \right]_{,j} \delta u_i + \left( \tilde{\mu}_{kij,k} + \lambda_{ij} \right) \delta e_{ij} \right\} dV \tag{80}$$

Therefore, the virtual work principle (75) becomes

$$\int_S \left[ t_i^{(n)} \delta u_i + \tilde{m}_{ij}^{(n)} \delta e_{ij} \right] dS + \int_V f_i \delta u_i dV = \int_S \left\{ \left[ \sigma'_{(ji)} + \lambda_{ij} \right] n_j \delta u_i + \tilde{\mu}_{kij} n_k \delta e_{ij} \right\} dS$$
$$- \int_V \left\{ \left[ \sigma'_{(ji)} + \lambda_{ij} \right]_{,j} \delta u_i + \left( \tilde{\mu}_{kij,k} + \lambda_{ij} \right) \delta e_{ij} \right\} dV \tag{81}$$
$$+ \int_V \rho a_i \, \delta u_i \, dV$$

Now by noticing that the variations $\delta u_i$ and $\delta e_{ij}$ are independent in the domain, we obtain the governing equations of motion as

$$\left[ \sigma'_{(ji)} + \lambda_{ij} \right]_{,j} + f_i = \rho a_i \tag{82}$$

$$\tilde{\mu}_{kij,k} + \lambda_{ij} = 0 \tag{83}$$

and the tractions as

$$t_i^{(n)} = \left[ \sigma'_{(ji)} + \lambda_{ij} \right] n_j \tag{84}$$

$$\tilde{m}_{ij}^{(n)} = \tilde{\mu}_{kij} n_k \tag{85}$$

Interestingly, by inspecting (82) and (83), we realize that the Lagrange multiplier tensor $\lambda_{ij}$ represents the second symmetric force-stress tensor part $\sigma''_{ij}$, where

$$\sigma''_{ij} = \lambda_{ij} = -\tilde{\mu}_{kij,k} \qquad (86)$$

Therefore, $\sigma_{ij} = \sigma'_{ij} + \sigma''_{ij}$ and the governing equations become

$$\sigma_{ji,j} + f_i = \rho a_i \qquad (87)$$

$$\tilde{\mu}_{kij,k} + \sigma''_{ji} = 0 \qquad (88)$$

We notice that the Lagrange multiplier virtual work principle gives both the force and first symmetric moment equations (87) and (88), similar to (35) and (36) for the systematic development in Section 3. However, in F-SGT, the force stress tensor $\sigma_{ij}$ and its decomposed parts $\sigma'_{ij}$ and $\sigma''_{ij}$ are always symmetric. The governing equations of motion (87) and (88) correspond to the translation $u_i$ and invalid strain degrees of freedom $e_{ij}$, respectively, where

$u_i \rightarrow$ 3 force equations $\qquad\qquad\qquad \sigma_{ji,j} + f_i = \rho \ddot{u}_i \qquad (89)$

$\varpi_{ij} = e_{ij} \rightarrow$ 6 non-physical symmetric moment equations $\quad \tilde{\mu}_{kij,k} + \sigma''_{ij} = 0 \qquad (90)$

The inconsistent governing equations (89) and (90) have been clearly presented in the developments by Mindlin and Eshel (1968); Georgiadis et al. (2000); Aifantis and Willis (2006); Aifantis et al. (2006); Fleck and Willis (2009, 2015); Niordson and Hutchinson (2011); Gudmundson (2004); Fredriksson et al. (2009); Fleck et al. (2014, 2015); and Lubarda (2016, 2017) in different formats.

### *4.2. Boundary conditions in the first strain gradient theory*

Since in F-SGT the tractions $t_i^{(n)}$ and $\tilde{m}_{ij}^{(n)}$ are energetically conjugate to $u_i$ and $e_{ij}$, the boundary conditions on the surface of the body can be either $u_i$ and $e_{ij}$ as essential (geometrical) boundary conditions, or $t_i^{(n)}$ and $\tilde{m}_{ij}^{(n)}$ as natural (mechanical) boundary conditions. Therefore, for boundary conditions one may specify displacements $u_i$ or force-tractions $t_i$

$$u_i = \bar{u}_i \quad \text{on } S_u \tag{91a}$$

$$t_i = \bar{t}_i \quad \text{on } S_t \tag{91b}$$

and strains $e_{ij}$ or double-tractions $\tilde{m}_{ij}$

$$e_{ij} = \bar{e}_{ij} \quad \text{on } S_e \tag{92a}$$

$$\tilde{m}_{ij} = \bar{\tilde{m}}_{ij} \quad \text{on } S_{\tilde{m}} \tag{92b}$$

Here $S_u$ and $S_e$ are the portions of the surface at which the essential boundary values for the displacement vector $u_i$ and the strain tensor $e_{ij}$ are prescribed, respectively. Furthermore, $S_t$ and $S_{\tilde{m}}$ are the portions of the surface at which the force-traction vector $t_i$ and the first symmetric double-traction tensor $\tilde{m}_{ij}$ are specified, respectively.

Accordingly, there are apparently a total number of nine boundary values for either essential or natural boundary conditions at each boundary point, respectively. However, we notice that if the components of $u_i$ are specified on the boundary surface, then the three in-plane strain components on the boundary surface are obtained from these boundary $u_i$, and cannot be prescribed independently. As a result, the total number of geometric or essential boundary conditions that can be specified on a smooth surface is six. This apparently shows that a material under F-SGT does not support independent distributions of in-plane double-traction components on the boundary surface, and the number of mechanical or natural boundary conditions also is six. However, the possibility of prescribing the strain $e_{ij}$ or first symmetric double-traction $\tilde{m}_{ij}$ on the boundary has no physical grounds, which again clearly shows the inconsistency of F-SGT.

Prescribing strains $e_{ij}$ as the essential or geometrical boundary conditions must have seemed very awkward to researchers in F-SGT. As a result, they did not consider its energy conjugate double-traction tensor $\tilde{m}_{ij}$ as the natural or mechanical boundary condition. Consequently, the proponents of F-SGT tried to find the apparently correct independent boundary conditions by transforming the tensorial boundary conditions $e_{ij}$ and $\tilde{m}_{ij}$ into some vectorial equivalent quantities. This can be seen in the developments by Mindlin and Eshel (1968), Fleck and Hutchinson (1997), Aravas (2011) and Gao and Park (2007), which we examine next. For this purpose, the external virtual work has been transformed to (Gao and Park, 2007)

$$\delta W_{ext} = \int_S \left[ \tilde{t}_i^{(n)} \delta u_i + q_i \delta \upsilon_i \right] dS + \int_C \varepsilon_{mlj} n_l \tilde{m}_{ij}^{(n)} \delta u_i dx_m + \int_V f_i \delta u_i dV \tag{93}$$

Here $\tilde{t}_i^{(n)}$ is the effective force-traction vector, $q_i$ is the reduced double-traction vector, and $\upsilon_i$ is the normal displacement derivative vector, where

$$\tilde{t}_i^{(n)} = \sigma_{ji} n_j - \left( n_{k,j} \tilde{\mu}_{kij} + n_k \tilde{\mu}_{kij,j} - n_j n_l n_{k,l} \tilde{\mu}_{kij} - n_j n_l n_k \tilde{\mu}_{kij,l} \right) + n_{l,l} n_j n_k \tilde{\mu}_{kij} \tag{94}$$

$$q_i = n_j \tilde{m}_{ij}^{(n)} = \tilde{\mu}_{kij} n_k n_j \tag{95}$$

$$\upsilon_i = n_l u_{i,l} = \frac{\partial u_i}{\partial n} \tag{96}$$

Consequently, in F-SGT, we may apparently specify displacements $u_i$ or the reduced force-tractions $\tilde{t}_i^{(n)}$

$$u_i = \bar{u}_i \quad \text{on } S_u \tag{97a}$$

$$\tilde{t}_i^{(n)} = \bar{\tilde{t}}_i^{(n)} \quad \text{on } S_t \tag{97b}$$

and normal displacement derivatives $\upsilon_i = \dfrac{\partial u_i}{\partial n}$ or reduced double-tractions $q_i$

$$\upsilon_i = \bar{\upsilon}_i \quad \text{on } S_e \tag{98a}$$

$$q_i = \bar{q}_i \quad \text{on } S_{\tilde{m}} \tag{98b}$$

Since the first double-traction and double-stress tensors $\tilde{m}_{ij}^{(n)}$ and $\tilde{\mu}_{ijk}$ are non-physical concepts, the effective force-traction vector $\tilde{t}_i^{(n)}$ with the awkward definition (94) and the reduced double-traction vector $q_i$ in (95) are still non-physical quantities. Also, the normal displacement derivative $\upsilon_i = \dfrac{\partial u_i}{\partial n}$ cannot be an essential or geometric boundary condition, because it does not define a set of degrees of freedom to describe the motion of an infinitesimal element of matter. Therefore, there are no correct boundary conditions in this inconsistent theory with nine inconsistent governing equations.

In couple stress theory, the boundary conditions on the surface of the body can be either $u_i$ and $\omega_i$ as essential (geometrical) boundary conditions, or $t_i^{(n)}$ and $m_i^{(n)}$ as natural (mechanical) boundary conditions. Furthermore, Mindlin and Tiersten (1962) and Koiter (1964) correctly established that five geometrical and five mechanical boundary conditions could be specified on a smooth surface. The consistency of these boundary conditions has revealed the skew-symmetric character of the couple-stress tensor in the determinate consistent couple stress theory (C-CST) (Hadjesfandiari and Dargush, 2011). However, Neff et al. (2015) have claimed discovering for the first time the complete, consistent set of traction boundary conditions for CST. For this purpose, they have replaced the rotation $\omega_i$ boundary condition by the normal displacement derivative $\upsilon_i = \dfrac{\partial u_i}{\partial n}$ from strain gradient theory. By imposing incorrect boundary conditions in the indeterminate couple stress theory, Neff and his colleagues are desperately trying to make the couple stress theory (CST) as a special case of the apparently more general strain gradient theories (SGT). However, this effort is in vain. As we have demonstrated, CST with six governing equations is not a special case of the non-physical first strain gradient theory (F-SGT) with nine governing equations. It is also not known why the couple-stress tensor in the framework of this set of supposedly complete boundary conditions would still be indeterminate. In fact, the entire development by Neff and colleagues is completely non-physical.

## 5. Some applications of first strain gradient theory

First strain gradient theory has been used extensively to develop size-dependent theories in many disciplines. Here, we examine the character of these developments for elasticity, plasticity, piezoelectricity and flexoelectricity. We also briefly investigate some fundamental inconsistencies of the first distortion gradient theory (F-DGT).

### *5.1. First strain gradient elasticity*

For a linear elastic body in F-SGT, the elastic energy density $U = U\left(e_{ij}, \tilde{k}_{ijk}\right)$ can be written as

$$U = \frac{1}{2} A_{ijkl} e_{ij} e_{kl} + \frac{1}{2} B_{ijklmn} \tilde{k}_{ijk} \tilde{k}_{lmn} + C_{ijklm} e_{ij} \tilde{k}_{klm}$$
$$= \frac{1}{2} A_{ijkl} e_{ij} e_{kl} + \frac{1}{2} B_{ijklmn} e_{ij,k} e_{lm,n} + C_{ijklm} e_{ij} e_{kl,m} \tag{99}$$

The tensors $A_{ijkl}$, $B_{ijklmn}$ and $C_{ijklm}$ contain the elastic constitutive coefficients and are such that $U$ is positive definite. We notice the symmetry relations are

$$A_{ijkl} = A_{jikl} = A_{klij} \tag{100}$$

$$B_{ijklmn} = B_{jiklmn} = B_{lmnijk} \tag{101}$$

$$C_{ijklm} = C_{jiklm} = C_{ijlkm} \tag{102}$$

which show that for the most general case the number of distinct components for $A_{ijkl}$, $B_{ijklmn}$ and $C_{ijklm}$ are 21, 171 and 108, respectively. Therefore, the most general linear elastic anisotropic material in F-SGT elasticity is described by 300 independent constitutive coefficients.

As a result, constitutive relations become

$$\sigma'_{ij} = \frac{\partial U}{\partial e_{ij}} = A_{ijkl} e_{kl} + C_{ijklm} e_{kl,m} \tag{103}$$

$$\tilde{\mu}_{kij} = \frac{\partial U}{\partial \tilde{k}_{ijk}} = B_{ijklmn} e_{lm,n} + C_{lmijk} e_{lm} \tag{104}$$

For a homogeneous material, the total force stress tensor becomes

$$\sigma_{ij} = A_{ijkl}e_{kl} + C_{ijklm}e_{kl,m} - C_{lmijk}e_{lm,k} - B_{ijklmn}e_{lm,kn} \qquad (105)$$

which in terms of displacements can be written as

$$\sigma_{ij} = A_{ijkl}u_{k,l} + C_{ijklm}u_{k,lm} - C_{lmijk}u_{l,km} - B_{ijklmn}u_{l,kmn} \qquad (106)$$

By carrying this relation into the reduced linear governing equation (70), we obtain the displacement governing equation as

$$A_{ijkl}u_{k,lj} + C_{ijklm}u_{k,lmj} - C_{lmijk}u_{l,kmj} - B_{ijklmn}u_{l,kmnj} + f_i = \rho \ddot{u}_i \qquad (107)$$

The appearance of up to 300 elastic constants in the constitutive relations would be clearly daunting in terms of material characterization. This appears even more absurd, when we notice that the force-stress tensor $\sigma_{ij}$ in this theory is symmetric after all. In a recent paper, Auffray et al. (2018) have presented the complete symmetry classification and compact matrix representations for this inconsistent strain gradient elasticity.

For an isotropic linear elastic material, this elasticity based F-SGT requires three extra elastic constants (Mindlin and Eshel, 1968), which also is not attractive from a practical perspective. It is obvious that in the framework of F-SGT there is no consistent solution for pure torsion of an isotropic elastic circular bar in static or quasistatic cases. The inconsistent elastic solution derived by Lazopoulos and Lazopoulos (2012) predicts significant size effect, which contradicts the observed no size effect in recent experiments for pure torsion of micro-diameter copper wires (Lu and Song, 2011; Song and Lu, 2015; Hadjesfandiari and Dargush, 2016b).

The inconsistent F-SGT has been used extensively to study size-dependent elastostatics and elastodynamics (e.g., Fannjiang et al., 2002; Polyzos et al., 2003; Paulino et al., 2003; Chan et al., 2008; Karlis et al., 2008; Gourgiotis and Georgiadis, 2009; Auffray, 2015; Auffray et al., 2013, 2015, 2018; Rosi and Auffray, 2016; Beheshti, 2017; Kolo et al., 2017; Gourgiotis et al., 2018). However, because of the various inconsistencies already noted in F-SGT, these developments cannot describe the behavior of materials correctly.

Altan and Aifantis (1997) have proposed simplified first strain gradient theory (SF-SGT) for elastic bodies to a form that requires only one strain gradient elastic constant. In this theory, the elastic constants are such that

$$B_{ijklmn} = \ell^2 A_{ijlm} \delta_{kn} \tag{108}$$

$$C_{ijklm} = 0 \tag{109}$$

where $\ell$ is the single material length scale parameter in this SF-SGT elasticity. Therefore, for the total symmetric force-stress tensor, we have

$$\begin{aligned}\sigma_{ij} &= \sigma'_{ij} - \ell^2 \nabla^2 \sigma'_{ij} \\ &= A_{ijkl}\left(e_{kl} - \ell^2 \nabla^2 e_{kl}\right)\end{aligned} \tag{110}$$

For an isotropic material, where

$$A_{ijkl} = \lambda \delta_{ij}\delta_{kl} + \mu \delta_{ik}\delta_{jl} + \mu \delta_{il}\delta_{jk} \tag{111}$$

the total force-stress tensor becomes

$$\begin{aligned}\sigma_{ij} &= \left(1 - \ell^2 \nabla^2\right)\sigma'_{ij} \\ &= \left(1 - \ell^2 \nabla^2\right)\left(\lambda e_{kk}\delta_{ij} + 2\mu e_{ij}\right)\end{aligned} \tag{112}$$

and the governing equation of motion in terms displacement becomes

$$\left(1 - \ell^2 \nabla^2\right)\left[(\lambda + \mu) u_{k,ki} + \mu \nabla^2 u_i\right] + f_i = \rho \ddot{u}_i \tag{113}$$

Nevertheless, this simplified strain gradient theory cannot describe the torsion of circular bars, and bending of beams and plates properly. This can be seen in the inconsistent stress solutions for static pure torsion of bars, and pure bending of beams and plates deforming to circular arcs and ellipsoidal surfaces, respectively.

The inability of SF-SGT elasticity to describe pure torsion of a bar and pure bending of a beam in hindsight demonstrates the inconsistency of the general first strain gradient theory (F-SGT). When a special case of a theory cannot account for size-effect for such a fundamental deformation as pure bending, it cannot be expected to describe deformations that are more complicated. Thus, it is not surprising to see that SF-SGT elasticity predicts no stress singularity for some cracked bodies (Altan and Aifantis, 1997), and significant size effect for pressurized thick-walled isotropic elastic cylinders (Gao and Park, 2007; Kolo et al., 2017). Therefore, the developments to account for size effects in the elasto-plastic analysis of a thick-walled cylinder or spherical shell, or cylindrical or spherical cavity by including strain gradients (Gao, 2002, 2003a,b, 2006; Tsagrakis et al., 2004; Zhuang et al., 2018) are not physically acceptable.

## *5.2. First strain gradient plasticity*

The F-SGT has been used extensively to develop size-dependent plasticity theories. In the developments by Aifantis and Willis (2006); Aifantis et al. (2006); Fleck and Willis (2009, 2015); Gudmundson (2004); Fredriksson et al. (2009); Fleck et al. (2014, 2015); Lubarda (2016, 2017), the nine inconsistent governing equations (87) and (88) have been explicitly presented in different formats. We notice that the developments based on the first strain gradient plasticity to model crack propagation (Nielsen et al., 2012) and damage modelling (Putar et al., 2017) are similarly non-physical.

There are also some other gradient plasticity developments. For example, in first distortion gradient theory (F-DGT) (Fleck and Hutchinson, 1997; Gao et al., 1999; Huang et al., 2000; Jiang et al., 2001; Hwang et al., 2002; Bažant and Guo, 2002; Qiu et al., 2003; Wulfinghoff et al., 2015), the measures of deformation are $e_{ij}$ and $u_{i,jk}$. As a result, the degrees of freedom of an infinitesimal element of matter in F-DGT total 12; three corresponding to translations $u_i$ and nine corresponding to the fictitious degrees of freedom distortion $u_{i,j}$. This requires 12 governing equations of motion in F-DGT. It turns out that the governing equations in F-DGT are based on the force and the invalid and non-physical first general moment equations (Hadjesfandiari and Dargush, 2018)

$$\sum F_i^{ext} = \sum ma_i \tag{114}$$

$$\sum \widehat{M}_{ij}^{ext} = \sum ma_i x_j \tag{115}$$

where $\widehat{M}_{ij} = F_i x_j$ is the first general moment of a force $F_i$ at $x_i$ about the origin. We notice that the first general moment of internal forces $\sum \widehat{M}_{ij}^{int}$ has been ignored in (115), which violates Newton's third law of action and reaction (Hadjesfandiari and Dargush, 2018). We also notice that the general moment tensor of a couple consisting of $F_i$ and $-F_i$ depends on the position of origin, and cannot describe its physical effect properly. As a result, in F-DGT, the internal stresses are represented by force stress tensor $\sigma_{ij}$, and the new higher stress $\widehat{\mu}_{ijk} = \widehat{\mu}_{jik}$. It is obvious that the set of 12 scalar equations (114) and (115) cannot describe the motion of a rigid body correctly. The corresponding 12 inconsistent governing equations for F-DGT have been explicitly presented in Wulfinghoff et al. (2015), whereas Fleck and Hutchinson (1997); Huang et al. (2000); Jiang et al. (2001); Hwang et al. (2002); Bažant and Guo (2002) and Qiu et al. (2003) have presented only the reduced linear governing equations. All of these developments suffer from fundamental inconsistencies and are clearly non-physical.

### *5.3. Size-dependent piezoelectricity and flexoelectricity based on strain gradient theory*

The confusion in the higher order continuum mechanics theories have had dramatic consequences in the development of some size-dependent multi-physics phenomena, such as size-dependent piezoelectricity and flexoelectricity. In strain gradient based flexoelectricity theory, the electric polarization is assumed to be generated as the result of coupling of the electric polarization vector $P_i$ (or electric field $E_i = -\phi$) to the third order strain gradient tensor, $\tilde{k}_{ijk} = e_{ij,k}$ (Tagantsev, 1986, Maranganti et al., 2006; Eliseev et al., 2009). As we can see, this theory suffers from the inconsistencies inherited from F-SGT.

Interestingly, in a review paper, Yudin and Tagantsev (2013) have stated that despite the considerable theoretical and experimental studies of flexoelectricity, there exist many open issues related to a limited understanding of the physics of flexoelectricity. As shown here, the main theoretical issues of this flexoelectricity theory are that the strain gradient $\tilde{k}_{ijk} = e_{ij,k}$ with 18 independent components is not a measure of deformation, and the third order double-stress tensor $\tilde{\mu}_{ijk}$ has no physical meaning. Furthermore, in this strain gradient flexoelectricity, the force stress tensor $\sigma_{ij}$ is always symmetric and the vector moment equation has been replaced with the non-physical first symmetric moment equation, which violates Newton's third law of action and reaction. As a result, the mechanical governing equations total nine instead of six. However, some proponents of this theory are not aware of this disturbing fact, because the governing mechanical equations are usually presented in the reduced form.

In a more recent article, Yudin and Tagantsev (2016) have also included the gradient of polarization $P_{i,j}$ in their enthalpy density function. This means $P_i$ (or $E_i$) is at the same time a set of degrees of freedom and an electric effect measure. The appearance of the gradient of polarization $P_{i,j}$ cannot be justified physically, because this creates a new electrical balance law in violation of Maxwell's equations (Hadjesfandiari, 2013, 2014). Therefore, this strain gradient flexoelectricity is inconsistent not only mechanically, but also electrically. We should notice that the inconsistent new electrical balance law is actually due to Mindlin (1968), who included $P_{i,j}$ in the enthalpy density function. Interestingly, he did not include any form of second gradient of deformation in his formulation. This clearly shows that he was not certain about the validity of any of his theories at the time to develop a size-dependent multi-physics formulation, such as size-dependent piezoelectricity or flexoelectricity. The similar inconsistency can also be seen in the formulation developed by Hu and Shen, (2009), where the gradient of electric field $E_{i,j}$ has been included in the enthalpy density function.

Inconsistencies of the strain gradient based size-dependent piezoelectricity show themselves in requiring too many elastic and flexoelectric coefficients, which also is not helpful from a practical perspective. It should be noticed that for isotropic or centrosymmetric cubic materials, this theory predicts three extra elastic coefficients and three flexoelectric coefficients. However, we should emphasize that the inconsistencies of this strain gradient based piezoelectricity are the main reason for rejecting it as a valid theory, not the appearance of too many constitutive coefficients.

We notice that the developments for flexoelectricity based strain gradient or higher gradient theory, such as Mao et al. (2014, 2016), Sladek et al. (2017, 2018), Deng et al. (2017, 2018) and Nanthakumar et al. (2017), do not describe physical problems correctly. It is surprising to see that Mao et al. (2014, 2016) predict a size effect and flexoelectric effect for a pressurized long thick-walled isotropic elastic cylinder. However, the size and flexoelectric-effects should be the result of flexure or bending deformation, as the term "flexoelectric" infers, not radial deformation.

## 6. Discussion

Here we have demonstrated the physical and mathematical inconsistencies of strain and distortion gradient theories (SGT and DGT). Since these theories are usually developed by using variational methods or a virtual work principle, its proponents do not realize that the underlying governing equations in this theory are invalid. For example, in F-SGT, the governing equations (54) and (55) are implicitly based on the force and first symmetric moment equations

$$\sum F_i^{ext} = \sum ma_i \qquad (116)$$

$$\sum \tilde{M}_{ij}^{ext} = \sum \frac{1}{2}\left(ma_i x_j + ma_j x_i\right) \qquad (117)$$

In particular, the first symmetric moment equation (117) is non-physical because the non-zero resultant internal moment $\sum \tilde{M}_{ij}^{int}$ has been neglected without any justification in this equation. This violates Newton's third law of action and reaction.

We have demonstrated that F-SGT is not based on the rigid body portion of motion of infinitesimal elements of matter at each point of the continuum. In this theory, the rotational degrees of freedom $\omega_i$ (or $\omega_{ij}$) have been replaced with the strain tensor $\varpi_{ij} = e_{ij}$ as the new degrees of freedom. This means that F-SGT cannot even describe the infinitesimal motion of a rigid body, let alone its size-dependent deformation. We also notice that in this theory, the fictitious degrees of freedom $\varpi_{ij} = e_{ij}$ can be specified on the boundary as an essential boundary condition similar to the displacement degrees of freedom $u_i$. Again, this is not physical.

It is interesting to notice that Lazopoulos (2009) and Münch et al. (2015) recognize the inconsistency of the artificial moment of couples equation in modified couple stress theory (M-CST) (Yang et al., 2002). However, they have not realized the inconsistency of the governing moment equation in gradient theories, such as (117) in the first strain gradient theory (F-SGT).

The main lesson is that a size-dependent elasticity theory cannot be developed by assuming an elastic energy density function based on some arbitrary measures of deformation, and then letting a variational method decide about the governing equations, boundary conditions and constitutive relations. The degrees of freedom, governing equations and the general form of tractions and stresses, though not in a final form, have already been known from the original developments of Mindlin and Tiersten (1962) and Koiter (1964) in couple stress theory, and these are consistent with the fundamentals of mechanics in describing the motion or equilibrium of infinitesimal elements of matter. Consequently, the results from variational methods for F-SGT are just mathematical developments without any connection to physical reality.

We have demonstrated that the first strain gradient theory (F-SGT) has no consistent relation with the original developments of Cosserat and Cosserat (1909), Mindlin and Tiersten (1962) and Koiter (1964) in couple stress theory, and suffers from the following inconsistencies:

1. This theory is not based on the rigid body portion of motion of infinitesimal elements of matter at each point of the continuum.

2. F-SGT cannot describe rigid body motion,

3. The number of degrees of freedom for infinitesimal elements of matter is nine; including three corresponding to translations $u_i$ and six corresponding to the incorrect strain degrees of freedom $e_{ij}$.
4. F-SGT violates Newton's third law of action and reaction.
5. This theory violates the angular momentum theorem.
6. The definition of the symmetric couple-traction $\tilde{m}_{ij}^{(n)}$ is based on the non-physical symmetric moment tensor $\tilde{M}_{ij} = \frac{1}{2}(F_i x_j + F_j x_i)$, rather than on the physical skew-symmetric moment $M_{ij} = \frac{1}{2}(F_i x_j - F_j x_i) = -M_{ji}$ or its dual pseudo-vector $M_i = \varepsilon_{ijk} x_j F_k$ ($\mathbf{M} = \mathbf{r} \times \mathbf{F}$).
7. The third order symmetric double-stress tensor $\tilde{\mu}_{ijk} = \tilde{\mu}_{ikj}$ has no physical meaning and cannot describe the internal stresses correctly.
8. There are nine equations of motion; including three based on the vectorial force and six based on the non-physical tensorial first symmetric moment governing equations.
9. Based on Noether's theorem (Noether, 1918), there is no symmetry of space corresponding to the first symmetric moment governing equations.
10. The strain tensor $e_{ij}$ is a measure of deformation and a set of degrees of freedom.
11. The first strain gradient $\tilde{k}_{ijk} = e_{ij,k}$ is a measure of deformation.
12. The strain degrees of freedom $e_{ij}$ can be defined as a set of boundary conditions.
13. The force-stress tensor $\sigma_{ij}$ is enforced arbitrarily to be symmetric.
14. There are up to 300 elastic coefficients in the constitutive relations for elastic bodies. For an isotropic material, these reduce to six elastic coefficients.
15. Simplified SF-SGT cannot describe the size-effect in beams and plates under static pure bending deformation, and cannot show the singularity of stresses in cracked bodies.

We notice that in consistent couple stress theory (C-CST), the mean curvature vector $\kappa_i$ with three components can be represented by a combination of 18 components of the first strain gradient tensor as $\kappa_i = \frac{1}{2}\left(e_{kk,i} - e_{ik,k}\right)$ (Hadjesfandiari and Dargush, 2015a,b). This means that although the first strain gradient tensor $\tilde{k}_{ijk} = e_{ij,k}$ is not itself a suitable measure of deformation, it still may appear in the analysis of deformation.

Researchers have also developed higher order gradient theories, which require introducing further non-physical higher order stresses and higher moment equations. By examining the second order strain gradient theory (S-SGT) (Mindlin, 1965; Polizzotto, 2003; Lazar et al., 2006; Cordero et al., 2016), we notice that the number of degrees of freedom for infinitesimal elements of matter is 27, with three translation degrees of freedom $u_i$, six fictitious strain degrees of freedom $e_{ij}$, and 18 fictitious strain gradient degrees of freedom $\tilde{k}_{ijk} = e_{ij,k}$. In this theory, the strain tensor $e_{ij}$, the strain gradient tensor $\tilde{k}_{ijk} = e_{ij,k}$ and the second strain gradient tensor $\tilde{k}_{ijkl} = e_{ij,kl}$ are measures of deformation. However, we notice that $e_{ij}$ and $\tilde{k}_{ijk} = e_{ij,k}$ are measures of deformation and degrees of freedom at the same time. As a result, in S-SGT, the internal stresses are represented by force stress tensor $\sigma_{ij}$, and the higher stresses $\tilde{\mu}_{ijk}$ and $\tilde{\mu}_{ijkl}$. We notice that the total force-stress tensor in S-SGT is symmetric, that is, $\sigma_{ij} = \sigma_{ji}$. The governing equations in this theory are based on the force and the non-physical first and second symmetric moment equations (Hadjesfandiari and Dargush, 2018):

$$\sum F_i^{ext} = \sum ma_i \tag{118}$$

$$\sum \tilde{M}_{ij}^{ext} = \sum \frac{1}{2}\left(ma_i x_j + ma_j x_i\right) \tag{119}$$

$$\sum \tilde{M}_{ijk}^{ext} = \sum \frac{1}{2}\left(ma_i x_j + ma_j x_i\right)x_k \tag{120}$$

However, we notice that the internal symmetric moments $\sum \tilde{M}_{ij}^{int}$ and $\sum \tilde{M}_{ijk}^{int}$ have been ignored in writing (119) and (120) by violating Newton's third law of action and reaction (Hadjesfandiari and Dargush, 2018).

For isotropic linear elastic material, S-SGT requires six strain gradient elastic coefficients (Mindlin and Eshel, 1968). Interestingly, Lazar et al. (2006) have simplified this theory further to a form, in which it requires only two strain gradient elastic coefficients. However, this simplified strain gradient theory cannot describe bending deformation properly, because it predicts inconsistent stresses for static pure bending of beams and plates. This inconsistency can be seen clearly for plate bending in Mousavi and Paavola (2014), which in hindsight demonstrates the inconsistency of second strain gradient theory (S-SGT).

It is more instructive, if we also examine the character of the second distortion gradient theory (S-DGT) of Mindlin (1965) and Polizzotto, (2016), where the measures of deformation are $e_{ij}$, $u_{i,jk}$, $u_{i,jkl}$. Consequently, in this inconsistent theory, the number of degrees of freedom for infinitesimal elements of matter is 30; three translation degrees of freedom $u_i$, nine fictitious distortion degrees of freedom $u_{i,j}$, and 18 distortion gradient fictitious degrees of freedom $u_{i,jk}$. This means the number of governing equations in this inconsistent theory is 30. As a result, in S-DGT, the internal stresses are represented by force stress tensor $\sigma_{ij}$, and the higher stresses $\hat{\mu}_{ijk}$ and $\hat{\mu}_{ijkl}$. Interestingly, for the sake of symmetry in the structure of S-DGT, one can argue that the measures of deformation should be $u_{i,j}$, $u_{i,jk}$, $u_{i,jkl}$. However, S-DGT with $u_{i,j}$ as a measure of deformation cannot recover the classical theory. In any case, following Noether's theorem (Noether, 1918), there would need to be 30 independent symmetries of space for S-DGT to be valid.

Consequently, higher order theories, such as S-SGT and S-DGT, suffer from many inconsistencies. These theories violate Newton's third law of action and reaction and the angular momentum theorem to create non-physical higher moment equations. **Table 1** summarizes the character of the different couple stress theories, and currently used strain and distortion gradient theories. This table presents the number of degrees of freedom, measures of deformation, stresses and governing equations in these theories.

**Table 1. Characteristics of different higher order theories**

| Theory | Degrees of freedom | Measures of deformation | Stresses | Number of governing equations |
|---|---|---|---|---|
| C-CST (Hadjesfandiari and Dargush, 2011) | 6<br>$u_i$, $\omega_i$ | 9<br>$e_{ij}$, $\kappa_i$ | 12<br>$\sigma_{ij}$, $\mu_i$ | 6 |
| M-CST (Yang et al., 2002) | 6<br>$u_i$, $\omega_i$ | 12<br>$e_{ij}$, $\chi_{ij}$ | 15<br>$\sigma_{ij}$, $\mu_{ij}$ | 9 |
| MTK-CST (Mindlin and Tiersten, 1962) | 6<br>$u_i$, $\omega_i$ | 15<br>$e_{ij}$, $\omega_{j,i}$ | 18<br>$\sigma_{ij}$, $\mu_{ij}$ | 6 |
| F-SGT First strain gradient theory | 9<br>$u_i$, $e_{ij}$ | 24<br>$e_{ij}$, $e_{ij,k}$ | 24<br>$\sigma_{ij}$, $\tilde{\mu}_{ijk}$ | 9 |
| F-DGT First distortion gradient theory | 12<br>$u_i$, $u_{i,j}$ | 24<br>$e_{ij}$, $u_{i,jk}$ | 24<br>$\sigma_{ij}$, $\hat{\mu}_{ijk}$ | 12 |
| S-SGT Second strain gradient theory | 27<br>$u_i$, $e_{ij}$, $e_{ij,k}$ | 60<br>$e_{ij}$, $e_{ij,k}$, $e_{ij,kl}$ | 60<br>$\sigma_{ij}$, $\tilde{\mu}_{ijk}$, $\tilde{\mu}_{ijkl}$ | 27 |
| S-DGT Second distortion gradient theory | 30<br>$u_i$, $u_{i,j}$, $u_{i,jk}$ | 51<br>$e_{ij}$, $u_{i,jk}$, $u_{i,jkl}$ | 51<br>$\sigma_{ij}$, $\hat{\mu}_{ijk}$, $\hat{\mu}_{ijkl}$ | 30 |

We have demonstrated that developing higher order theories is not that arbitrary, because any higher strain and distortion gradient measure of deformation requires a new set of non-physical governing moment equations and inconsistent boundary conditions. Therefore, contrary to the claim of Neff et al. (2016), couple stress theory (CST) with six governing equations is not a special case of general non-physical strain gradient theories with nine or more non-physical governing equations. Furthermore, it is very strange to see that after half a century, Neff and his colleagues claim that there is no problem in an indeterminate couple stress theory (Neff et al., 2016; Ghiba et al., 2016; Madeo et al., 2016; Münch et al., 2015), and also think strain gradient theories are valid. In the misleading titled paper, Neff et al. (2016) claim that the couple-stress tensor can be taken deviatoric, which is physically impossible (Hadjesfandiari and Dargush, 2015b, 2016a). Forgetting to impose the compatibility of rotation vector in their variational method, Neff and his colleagues (Neff et al., 2016; Ghiba et al., 2016; Madeo et al., 2016; Münch et al., 2015) and Park and Gao (2008) have incorrectly concluded that a direct variational method for linear elastic bodies results in a trace free couple-stress tensor (Hadjesfandiari and Dargush, 2016a). As discussed in Section 4, these authors have also incorrectly claimed discovering the correct traction boundary conditions in the indeterminate couple stress theory (Neff et al., 2015).

There has been some effort to develop higher gradient formulations for large deformation problems, such as finite elasticity (e.g., dell'Isola et al., 2009; dell'Isola and Steigmann, 2015; Ferretti et al., 2014; Madeo et al., 2015; Barbagallo et al., 2017a,b). Naturally, these formulations are inconsistent and suffer from similar physical and mathematical inconsistencies. It turns out that the number of degrees of freedom for infinitesimal elements of matter in these theories are 12 or 9, which requires 12 or 9 governing equations of motion. Therefore, these developments cannot describe large deformations correctly. It is logical to finalize the small deformation size-dependent theory first, and then develop the corresponding large deformation theory.

## 7. Conclusions

In this paper, we have examined the physical and mathematical inconsistencies of higher gradient theories such as F-SGT, D-SGT, S-SGT and S-DGT. Although the deformation in these theories is completely specified by the continuous displacement field $u_i$, the degrees of freedom describing the motion of infinitesimal of elements of matter are not correct. We notice that the displacement field $u_i$ is the translation degrees of freedom at each point, but there is no explicit rotation degrees of freedom $\omega_i$ in these theories. As a result, the measures of deformation, internal stresses and the governing equations are inconsistent in these higher gradient theories. The moment governing equations in these theories are based on non-physical equations, which violate Newton's third law of action and reaction and the angular momentum theorem. Therefore, the higher gradient theories are not an extension of rigid body mechanics, which then cannot be recovered in the absence of deformation.

Since the higher gradient theories are usually developed by using variational methods or a virtual work principle, the existence of these many inconsistencies, such as nine or more non-physical governing equations of motion, have not been recognized previously. We notice that variational methods are very efficient, but are only useful when the correct degrees of freedom and measures of deformation are already known. Therefore, higher gradient theories are just mathematical developments, without any physical reality. This is the power of mathematics that although higher gradient theories are incorrect, variational methods still result in a beautiful mathematical, but totally non-physical, set of governing equations, constitutive relations and boundary conditions. Therefore, variational methods and virtual work methods can be very misleading tools in developing continuum mechanics, when the degrees of freedom and measures of deformation are not known beforehand. This indicates that changes in mechanics education also are needed to avoid future missteps along non-physical directions.

Fortunately, consistent couple stress theory (C-CST) has been established by discovering the skew-symmetric character of the couple-stress tensor. It turns out that C-CST provides a fundamental basis for the development of many linear and non-linear size-dependent multi-physics phenomena in continuum mechanics.

Green, A.E., Rivlin, R.S., 1964a. Simple force and stress multipoles, Arch. Rat. Mech. Anal. 16, 325–353.

Green, A.E., Rivlin, R.S., 1964b. Multipolar continuum mechanics. Arch. Rat. Mech. Anal. 17, 113–147.

Gudmundson, P., 2004. A unified treatment of strain gradient plasticity. J. Mech. Phys. Solids, 52(6), 1379-1406.

Hadjesfandiari, A. R., 2013. Size-dependent piezoelectricity. Int. J. Solids Struct. 50, 2781-2791.

Hadjesfandiari, A. R., 2014. Size-dependent theories of piezoelectricity: Comparisons and further developments for centrosymmetric dielectrics. Preprint arXiv: 1409.1082.

Hadjesfandiari, A. R., Dargush, G. F., 2011. Couple stress theory for solids. Int. J. Solids Struct. 48 (18), 2496-2510.

Hadjesfandiari, A. R., Dargush, G. F., 2015a. Evolution of generalized couple-stress continuum theories: a critical analysis. Preprint arXiv: 1501.03112.

Hadjesfandiari, A. R., Dargush, G. F., 2015b. Foundations of consistent couple stress theory. Preprint arXiv: 1509.06299.

Hadjesfandiari, A. R., Dargush, G. F., 2016a. Couple stress theories: Theoretical underpinnings and practical aspects from a new energy perspective. Preprint arXiv: 1611.10249.

Hadjesfandiari, A. R., Dargush, G. F., 2016b. Comparison of theoretical elastic couple stress predictions with physical experiments for pure torsion. Preprint arXiv: 1605.02556.

Hadjesfandiari, A. R., Dargush, G. F., 2018. Fundamental governing equations of motion in consistent continuum mechanics. Preprint arXiv: 1810.04514.

Hadjesfandiari, A. R., Hajesfandiari, A., Dargush, G. F., 2015. Skew-symmetric couple-stress fluid mechanics. Acta Mech. 226 (2015) 871–895.

Hu, S.L., Shen, S.P., 2009. Electric field gradient theory with surface effect for nano-dielectrics. CMC 13(1), 63–87.

Huang, Y., Gao, H., Nix, W.D., Hutchinson, J.W., 2000. Mechanism-based strain gradient plasticity—II. Analysis. J. Mech. Phys. Solids, 48(1), 99-128.

Hwang, K.C., Jiang, H., Huang, Y., Gao, H. and Hu, N., 2002. A finite deformation theory of strain gradient plasticity. J. Mech. Phys. Solids, 50(1), 81-99.

Jiang, H., Huang, Y., Zhuang, Z. and Hwang, K.C., 2001. Fracture in mechanism-based strain gradient plasticity. J. Mech. Phys. Solids, 49(5), 979-993.
42